# Upgrade of the ALICE Time Projection Chamber for the LHC Run3

**A.L.Gera[1], for the ALICE TPC collaboration**
*Wigner Research Centre for Physics*
*29-33 Konkoly-Thege Miklós str, 1121 Budapest, Hungary*
*E-mail:* gera.adam@wigner.mta.hu

**Abstract**

ALICE, one of the experiments at the Large Hadron Collider (CERN), is undergoing ambitious upgrades during the ongoing second long shutdown (LS2). The main goal of this project is to access rare events and previously inaccessible physics observables. The increased Pb–Pb and pp collision rates need a correspondingly higher TPC operation rate in the next Run 3 of the LHC.

The operational gated MWPC readout used so far cannot be used at such fast rates. Therefore a new readout chamber is needed with a novel technology and without any compromise on the momentum and energy resolution. As a solution the Gas Electron Multiplier (GEM) technology is applied. This new readout chamber consists of inner- (IROC) and outer (OROC) readout chambers with a 4 stage GEM cascade in order to reduce back-drifting ion space charges. These quadruple stacks have proven to provide sufficient ion blocking capabilities. This structure also preserves the intrinsic energy resolution and keeps the space-charge distortions at a tolerable level. The GEMs cannot be repaired during operation, therefore to minimize the unsuitable foils, the Quality Assurance system is developed. These cleanroom measurements investigate the hole size uniformity, the gain uniformity and the electrical stability. Due to the continuous readout, higher readout rate is possible. The new ASIC-SAMPA has been developed for this purpose.

After the ROC assembly some of chambers were tested in the ALICE cavern, and a test campaign started for the further testing of the ROC bodies at GIF[++] at CERN.

---

[1]Speaker

## 1.   Introduction

ALICE, one of the experiments at CERN Large Hadron Collider (LHC), focuses on the study the physics of strongly interacting matters, especially in ultra-relativistic central heavy-ion collision, where the quark-gluon plasma (QGP) forms. The ALICE detector [7] has demonstrated its excellent particle identification (PID) capabilities during Run 1 and Run 2 with an integrated luminosity of 1 $nb^{-1}$. After the ongoing Long Shutdown 2 (LS2) the luminosity of the LHC will be increased to 10 $nb^{-1}$ for heavy ions, with 50 kHz collision rates for Pb–Pb. The main goal is to access rare events and previously inaccessible physics observables which can be achieved through high precision and high statistics. The ALICE detector upgrade aims to be able to make full use of the increased LHC luminosity for Pb–Pb [1].

## 2.   The ALICE TPC and its upgrade

The ALICE Time Projection Chamber (TPC) is the main tracking and particle identification detector in the central barrel. A sketch of the TPC is shown in Fig. 1. The detector field cage is cylindrical in shape with an active volume of about 90 $m^3$. At both ends the endplates keep together the cylinders and also hold the readout chambers. The TPC used Ne-$CO_2$ (90-10) as gas mixture, but simulations have shown that the Ne can be replaced by Ar, and this gas mixture can also be used. The addition of 5 % $N_2$ was successfully tested and in Run 3 the gas mixture will be Ne-$CO_2$-$N_2$ (90-10-5) [2].

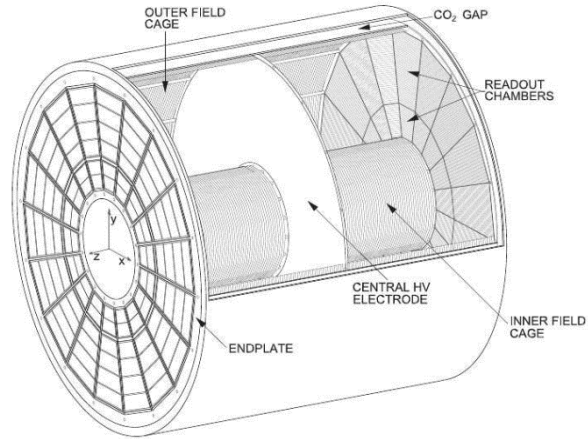

Figure 1: The ALICE Time Projection Chamber. The new readout chambers and electronics will replace the existing ones, while other parts and the gas system will remain the same.

The previous MWPC-based readout chambers used a plane of anode wires, cathode wires and a gating grid (GG) during Run 1 and Run 2. After a trigger, it allows the ionization electrons to pass into the active area. In order to provide efficient ion blocking, the gating grid must be closed for the maximum ion drift time (180 μs) with an alternating potential. Leaving the gating grid continuously open would result in a massive space charge distortion in the drift volume due to the ion backflow. On the other hand this alternating operation sets a limitation of about 3.5 kHz to the maximum readout. In the previous runs it matched well with the Pb–Pb collision rates. However in Run 3 with the increased luminosity the gating grid is not an option anymore and the TPC will have to operate continuously with new readout chambers that have intrinsic ion-blocking capabilities [1]. The field cage and most of the services of the previous TPC will remain the same, although to operate in an ungated mode the MWPC-based readout chambers will be replaced by Gas Electron Multipliers (GEM) in quadruple stacks [3].

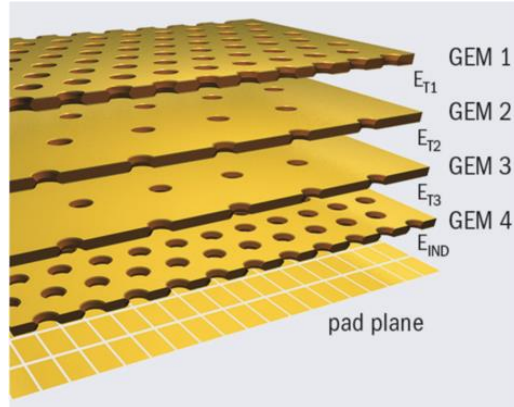

Figure 2: The schematic view of the quadruple GEM stack. The first and the last foils are standard pitch GEMs (140 μm pitch S), the middle foils are large pitch GEMs (280 μm pitch LP).

The result of a major R&D effort (Micro-Pattern Gas Detectors or MPGD) by the RD51 collaboration [8] was a genuinely new concept in the world of gaseous detectors. This development uses cutting edge technology to manufacture these kind of advanced gas-avalanche detectors, for instance the GEMs are chemically pierced with a high hole density to maintain the amplification. One of the most important advantages of this kind of detectors is that it can be cascaded, which is a key element in the ALICE TPC continuous readout. Several years of extensive research led to the quadruple GEM stacks, which have proven to provide sufficient ion blocking capabilities. The stacks will operate at a total effective gain of 2000. At this gain, to keep space-charge distortions at a tolerable level the fractional ion backflow must be below 1% while the ionization energy loss ($dE/dx$) and momentum resolution shall remain the same. Previous studies have shown that the quadruple stack with the proper layout as Fig. 2 shows, can be optimized with respect to the ion backflow and the required energy resolution. The result of the optimization process can be seen on Fig. 3.

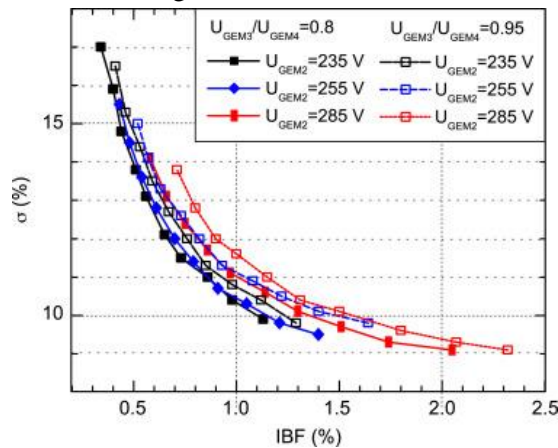

Figure 3: Correlation between ion backflow and energy resolution at 5.9 keV in a quadruple S-LP-LP-S GEM in Ne-$CO_2$-$N_2$ (90-10-5) gas for various voltage settings [2].

### 3. GEM Quality Assurance (QA)

The GEMs cannot be repaired during operation, therefore it was a priority to minimize the number of unsuitable foils. Due to the cutting edge technology, production issues can occur. We require the quality selection of the GEMs, which focused on choosing foils with the best gain uniformity, and long-term electrical stability. This operation was divided into basic QA and advanced QA where the measurements were made in certificated cleanrooms. During basic QA, at the production site (CERN), a short (30 min) leakage current measurement and an optical inspection were made. The advanced QA, which was done in two established QA centres in Helsinki and Budapest, included more detailed measurements as presented in Fig. 4 as an example. First an optical scanning was preformed to measure hole size uniformity and detect microscopic defects with an automated high resolution microscopic camera mounted onto an xyz-table. This was followed by a long-term high voltage test to check the electrical stability. Finally the gain uniformity was tested for selected foils, to validate the predicted gain map from the optical scanning. The two QA centers inspected more than 720 foils during the two years period [4].

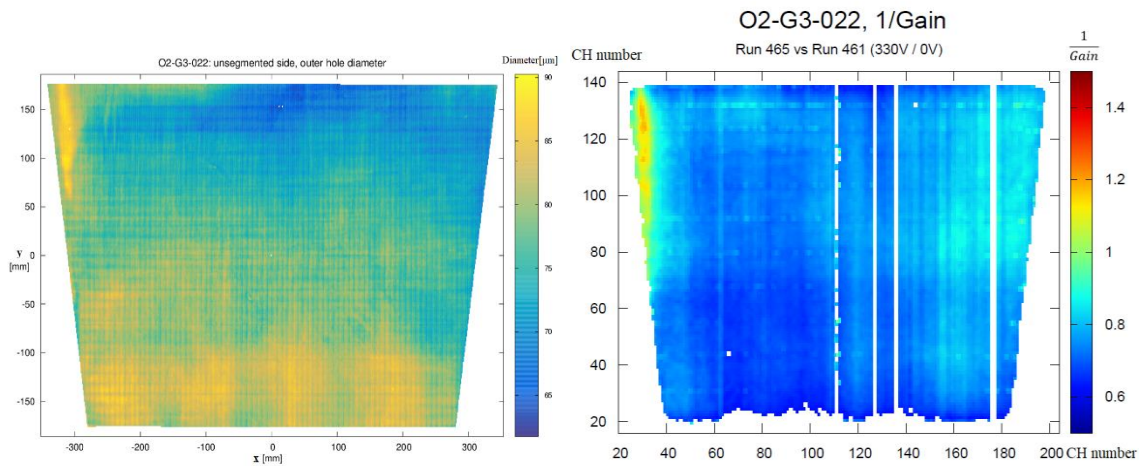

Figure 4: The optical scanning (left) and the gain measurement (right) of the same GEM foil. On the left figure the outer diameter map is shown in the *x-y* plane, on the right figure the reciprocal of the gain is shown in the channels. At the top left corner the same formation can be seen.

### 4. New front-end and readout electronics

The increased LHC luminosity implies that the drift time will be higher than the average time between interactions, therefore continuous readout were developed and implemented for Run 3. This means that higher readout rate is necessary and the existing front-end ASICs have to be upgraded as well. The new read-out SAMPA ASIC provides continuous read-out and also can handle different detector signal polarities. It will integrate 32 channels per chip which digitize and process the input signals. The schematic view of the new read-out can be seen on Fig. 5. The new read-out offers continuous sampling at 10 bit and 20 Msamples/s through electrical links. After that a digital signal processor is used to eliminate signal perturbations, moreover offset and signal variation can be applied to account for temperature changes [5].

The front-end cards and the online farm are connected via the Common Read-out Unit (CRU) and the expected data output for 50 kHz Pb–Pb collisions is up to 1 TB/s [6].

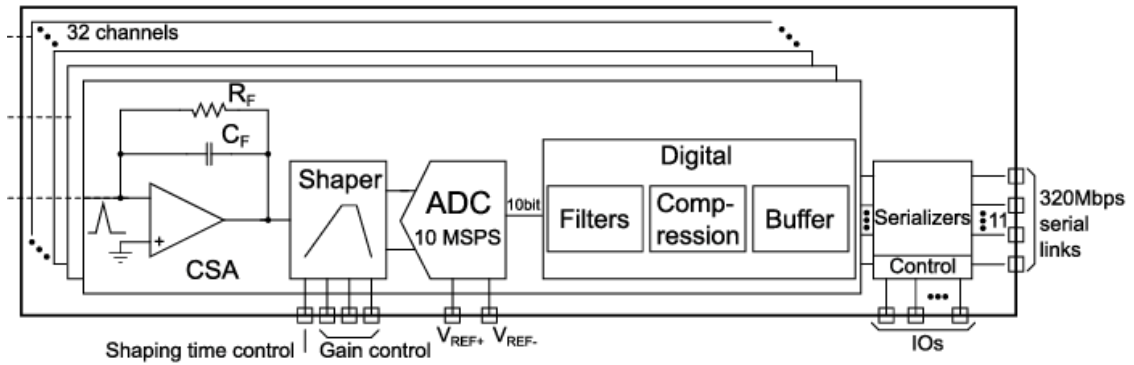

Figure 5: SAMPA system block diagram [5].

## 5. ROC tests in ALICE cavern and GIF[++]

The new readout chambers which have inner and outer section (IROC: one-GEM stack, OROC: three-GEM stack) are already finished and tested under radiation conditions that are comparable to Run 3 in the ALICE detector.

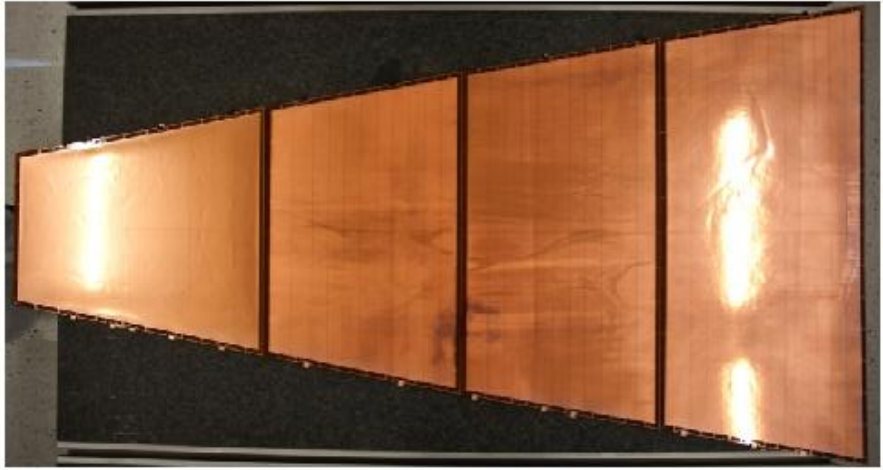

Figure 6: One full sector of the ALICE TPC [6].

Some of the IROCs and OROCs as seen on Fig. 6 were tested before LS2 in an area close to the LHC beam pipe which fullfils the Run 3 conditions, but not all of the chambers. A test campaign at the Gamma Irradiation Facility (GIF[++]) at CERN was done. The large radiation fields of GIF[++] allow for the simultaneous testing of different detectors with a 14 TBq $^{137}$Cesium source. The ALICE TPC collaboration managed to transport and test up to 8 Read Out Chambers (ROC) per week until the successful end of the campaign. This was also a good opportunity to recognize any kind of instabilities and repair these chambers. Currently the assembly of the ROCs is ongoing, including cabling and FEE installation.

## 6.     Summary

During the LS2 in 2018/2019 an extensive upgrade is ongoing related to the ALICE experiment. The increased LHC luminosity implies continuous readout of the TPC with low ion backflow and good resolution without the use of the gating grid. The solution is the quadruple stack GEM which was successfully optimized through major R&D process for the IBF and the d$E$/d$x$. This will be the largest GEM-based detector once it is in operation. After the production of the foils the quality assurance programme with its advanced measurements was successful and the ROC assembly could start. The testing of the assembled chambers was performed in the ALICE cavern and at GIF$^{++}$ where the conditions are comparable to Run 3. After the transportation of the TPC to the ground, cleaning and irradiation tests were done in the cleanroom. The new readout chambers are installed by now and the TPC will be ready for transportation at early 2020, as planned.

**Acknowledgement**

I appreciate the help of Gergely Gábor Barnaföldi and Dezső Varga, the ALICE Budapest group and the REGARD group. This work was supported by Hungarian NKFIH OTKA K120660 and Lendület grants. Computational resources were provided by the Wigner GPU Laboratory.